\documentstyle{elsartwb}
\input epsf.sty

\begin{document}
\newsavebox{\tabela}

\begin{frontmatter}
\title{Gradient terms in the microscopic description of K-atoms}
\author[Darmstadt]{Matthias Lutz} and \author[Krakow]{Wojciech Florkowski}
\address[Darmstadt]{Gesellschaft f\"ur Schwerionenforschung GSI,
Postfach 110552, D-64220 Darmstadt, Germany}
\address[Krakow]{H. Niewodnicza\'nski Institute of Nuclear Physics
ul. Radzikowskiego 152, PL-31-342 Krak\'ow, Poland}

\begin{abstract}
We analyze the spectra of kaonic atoms using optical potentials with
non-local (gradient) terms. The magnitude of the non-local terms
follows from a self consistent many-body calculation of the kaon self
energy in nuclear matter, which is based on s-wave kaon nucleon
interactions.  The optical potentials exhibit strong non-linearities
in the nucleon density and sizeable non-local terms. We find that the
non-local terms are quantitatively important in the analysis of the
spectra and that a phenomenologically successful description can be 
obtained for p-wave like optical potentials. It is suggested that the
microscopic form of the non-local interaction terms is obtained
systematically by means of a semi-classical expansion of the nucleus
structure. The resulting optical potential leads to less pronounced 
non-local effects.
\end{abstract}
\end{frontmatter}


\section{Introduction}

Kaonic atom data provide a valuable consistency check on any
microscopic theory of the $K^-$ nucleon interaction in nuclear
matter. We therefore apply the microscopic approach, developed by one
of the authors in \cite{ML}, to kaonic atoms. The description of the
nuclear level shifts in $K^-$ atoms has a long history. For a recent
review see \cite{report}. We recall here the most striking puzzle. In
the extreme low-density limit the nuclear part of the optical potential, 
$U_{\rm opt.}$, is determined by the s-wave $K^-N$ scattering length:
\begin{eqnarray}
2\,\mu\,\,U_{\rm opt.} \left(\vec r \,\right) =
-4\,\pi \left( 1+\frac{m_K}{m_N} \right) a_{KN}\,\rho (r)
\label{LDT}
\end{eqnarray}
with $ a_{KN} ={\textstyle{1\over 4}}\,a_{KN}^{(I=0)}+{\textstyle{3\over 4}}\,
\,a_{KN}^{(I=1)} \simeq \left( -0.18 +0.67\,i \right) \, \mbox{fm}$,
the nucleus density profile $\rho(r)$ and the reduced kaon mass $\mu$
of the $K^-$ nucleus system. As shown by Friedman, Gal and Batty
\cite{Friedman} kaonic atom data can be described with a large
attractive effective scattering length $a_{\rm eff.} \simeq
\left( 0.63 +0.89 \,i\right)  $ fm,  which is in direct 
contradiction to the low-density optical potential (\ref{LDT}). The
present data set on kaonic level shifts include typically the
3d$\to$2p transition for light nuclei and the 4f$\to$3d transition for
heavy nuclei. Deeply bound kaonic states in an s-wave have not been observed so
far.  Since a $K^-$ bound in a p-wave at a nucleus
probes dominantly the low-density tail of the nucleus profile one may
conclude that the optical potential must exhibit a strong non-linear
density characteristic at rather low density.

A further complication was pointed out for example by Thies \cite{Thies},
who demonstrated that in the kaonic $^{12}_6 C $ system
non-local effects may be important. The importance of non-local effects 
was also emphasized in \cite{Bardeen,Alberg}. In fact a reasonable description
of kaonic atom level shifts was achieved by Mizoguchi, Hirenzaki and
Toki \cite{Toki} with a phenomenological non-local optical potential
of the form
\begin{eqnarray}
2\,\mu\,\,U_{\rm opt.} \left(\vec r,\vec \nabla \right) =
-4\,\pi \left( 1+\frac{m_K}{m_N} \right)
\left(a_{KN}\,\rho (r)
- b\, \vec \nabla \,\rho(r)\cdot \vec \nabla \right) \;,
\label{potential2}
\end{eqnarray}
where the parameter $b \simeq (0.47 +i\,0.30 ) $ fm$^3$. 

Though it has been long anticipated that the $\Lambda (1405) $ 
resonance plays a key role \cite{Bardeen,Alberg,Toki,Brockmann}, a microscopic 
description of kaonic atom data remains a challenge. Obviously an 
important ingredient of any such attempt must be a proper many-body 
treatment of the $\Lambda (1405)$ resonance structure in nuclear 
matter. 

\section{ Kaon self energy in nuclear matter}

In this section we prepare the ground for our study of kaonic
atoms. Of central importance is the kaon self energy, $\Pi_{K}(\omega
,\vec q\, ; \rho )$, evaluated in nuclear matter. Here we introduce
the self energy relative to all vacuum polarization effects,
i.e. $\Pi_{K}(\omega ,{\vec q}\, ; 0 )= 0$.

First we recall results for the kaon self energy as obtained in a self
consistent many-body calculation based on microscopic s-wave
kaon-nucleon interactions.  For the details of this microscopic
approach we refer to \cite{ML}. We identify the $K^-$-nucleus optical
potential $V_{\rm opt.} (\vec q \,)$ by
\begin{eqnarray}
2\,E_K({\vec q}\,)\, V_{\rm opt.} ({\vec q}\, ) =\Pi_{K}( \omega 
=E_K({\vec q}\,), {\vec q}\, ) \;,
\end{eqnarray}
where $E_K({\vec q}\, )=(m_K^2+{\vec q}\,^2 )^{1/2}$ is the free-space kaon 
energy. In Fig. 1 we present the result  of \cite{ML} 
\begin{figure}[t]
\epsfysize=6.5cm
\begin{center}
\mbox{\epsfbox{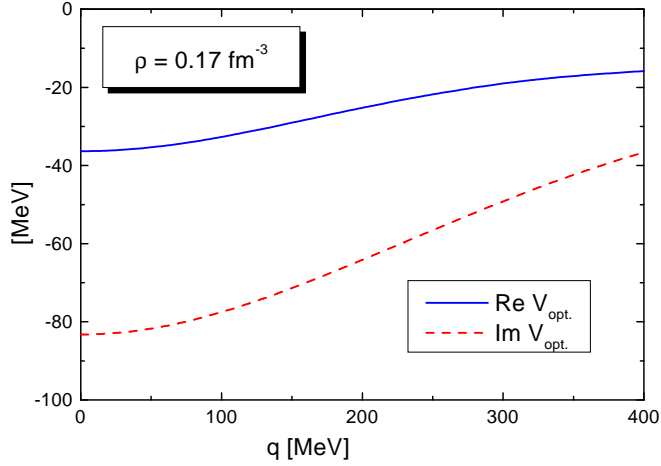}}
\end{center}
\caption{The $K^-$ nuclear optical potential, $V_{\rm opt.}(\vec q\,) $,
plotted as a function of the kaon three momentum. The nuclear 
density is $\rho =0.17  $ fm$^{-3}$.} 
\label{fig1}
\end{figure}
at nuclear saturation density. We point out that the real part of our
optical potential exhibits rather moderate attraction of less than 40
MeV. On the other hand we find a rather strong absorptive part of the
optical potential. This is in disagreement with recent work by
Friedman, Gal, Mares and Cieply \cite{Mares}.  We recall that in
\cite{ML} the quasi-particle energy for a $K^-$ at rest was found to
be $ 380 $ MeV at nuclear saturation density and that the large
attraction in the quasi-particle energy is consistent with the
moderate attraction in the optical potential.  It merely reflects the
strong energy dependence of the kaon self energy induced by the
$\Lambda (1405) $ resonance. Such important energy variations are
typically missed in a mean field approach. Hence, a proper treatment
of the pertinent many-body effects is mandatory.

We analyze the self energy as probed in kaonic atoms in more
detail. Before we proceed it is important to straighten out the
relevant scales. Since the typical binding energy of a $K^-$ bound at
a nucleus in a p-wave is of the order of 0.5 MeV the typical kaon
momentum is estimated to be roughly 20 MeV. We then expect the relevant
nuclear Fermi momentum $k_F$ to be larger than the kaon momentum
$|\vec q\,|< k_F $.
\begin{figure}[t]
\epsfysize=7.0cm
\begin{center}
\mbox{\epsfbox{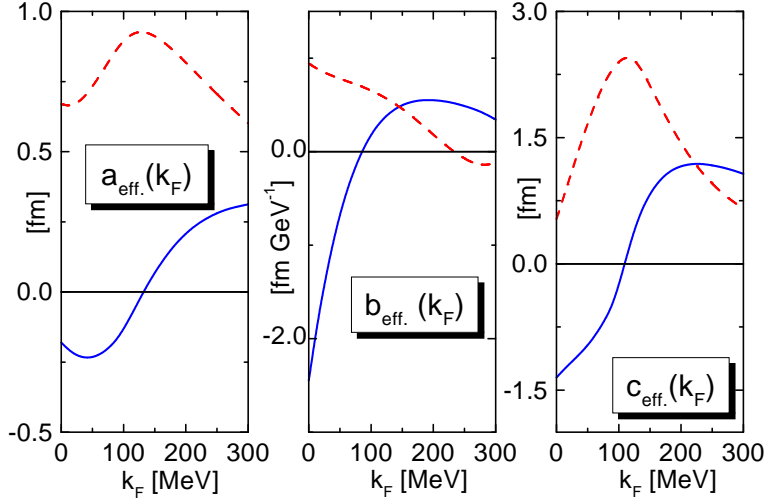}}
\end{center}
\caption{The effective scattering length $a_{\rm eff.}(k_F)$ and 
effective slope parameters $b_{\rm eff.}(k_F) $ and $c_{\rm eff.}(k_F)$ as 
defined in (\ref{beff}). The solid and dashed lines represent the 
real and imaginary parts, respectively. } 
\label{fig2}
\end{figure}
For the study of kaonic  level shifts it is therefore useful to 
introduce the effective scattering length $a_{\rm eff.}(k_F) $ and the 
effective slope parameters $b_{\rm eff.}(k_F)$ and $c_{\rm eff.}(k_F)$ 
\begin{eqnarray}
\Pi_{K} (\omega , {\vec q }\,)
&=&-\frac{8}{3\,\pi} \left(1+\frac{m_K}{m_N} \right)
\left( a_{\rm eff.}(k_F)\,k_F^3
+ b_{\rm eff.}(k_F)\,k^2_F\,{\vec q}\,^2 \right)
\nonumber\\
&+&\frac{8}{3\,\pi} \left(1+\frac{m_K}{m_N} \right)
c_{\rm eff.}(k_F)\,k^2_F\,\Big(\omega -m_K \Big)
\nonumber\\
&+&{\mathcal O}\left( {\vec q}\,^4 , \left(\omega -m_K \right)^2, {\vec 
q}\,^2 \, \left(\omega -m_K \right)\right), 
\label{beff}
\end{eqnarray}
where we expanded the kaon self energy for small momenta  $ {\vec
q} $ and energies close to $m_K$\footnote{We emphasize that
expressions (\ref{beff},\ref{self-expand}) do not contradict the
low-density theorem. Taking for instance the derivative of $\Pi (\omega , \vec q\,)$ 
with respect to $q$ assumes implicitly $q <k_F$ if the self energy 
shows a contribution depending on the ratio $q^2/k_F^2$. Pauli blocking 
does lead to such a contribution as shown in the Appendix. 
Consequently, our optical potential is necessarily incorrect for $k_F < q \sim 20$ MeV 
where we recall the typical 3-momentum of the kaon probed in K-atoms. 
We anticipate that the systematic error we encounter with 
our expansion in (4) is insignificant since the optical potential is 
already tiny by itself in the region $0 < k_F < 20$ MeV. The main contribution 
to the kaonic level shifts is expected for $k_F \gg q $.}. The rational 
behind the expansion (4) will be discussed in more detail in the subsequent 
section when constructing the optical potential. It is
instructive to derive the model independent low-density limit of the
slope parameters: 
\begin{eqnarray}
\!
b_{\rm eff.}(k_F=0)\!\! &&=\frac{1}{8 \,\pi}
\left( 
\frac{2}{1-\kappa}+\frac{7-3\,\kappa^2}{1-\kappa^2}\,
\frac{\log |\kappa |}{1-\kappa}
\right)
\Bigg(
\left(a_{KN}^{(I=0)}\right)^2 + 
3\left(a_{KN}^{(I=1)}\right)^2 \Bigg), 
\nonumber\\
\!
c_{\rm eff.}(k_F=0)\!\! &&= \frac{3\,m_K}{4\,\pi}\,
\frac{\log |\kappa | }{1-\kappa }\,
\Bigg(
\left(a_{KN}^{(I=0)}\right)^2 + 
3\left(a_{KN}^{(I=1)}\right)^2 \Bigg),
\label{self-expand}
\end{eqnarray}
which is given in terms of the kaon-nucleon scattering lengths and the
ratio $\kappa =m_K/m_N$. As demonstrated in the appendix these limits
are determined by the Pauli blocking effect which characterizes the
leading medium modification of the kaon-nucleon scattering
process. Using the values for the scattering lengths we obtain $b_{\rm
eff.}(0) \simeq  \left( -2.45+i\,0.95 \right)$ fm GeV$^{-1}$ and 
$c_{\rm eff.}(0) \simeq \left( -1.35+i\,0.53 \right)$ fm.

In Fig. 2 the effective scattering length and the slope parameters are
presented as extracted numerically from the self consistent
calculation of \cite{ML}.  The real part of the effective scattering
length $a_{\rm eff.}(k_F) $ changes sign as the density is
increased. At large densities we find an attractive effective
scattering length in qualitative agreement with the previous work
\cite{Waas}.  Note that the quantitative differences of the works
\cite{Waas} and \cite{ML} are most clearly seen in the imaginary part
of the effective scattering length in particular at small density. A
more recent calculation by Ramos and Oset \cite{Ramos} confirms the
results of \cite{ML} at the qualitative level. The quantitative
differences will be discussed in the next section. Most dramatic are
the non-linear density effects in the effective slope parameters
$b_{\rm eff.}(k_F)$ and $c_{\rm eff.}(k_F)$, not considered in
\cite{Waas,Koch,Ramos}. 

The non-trivial changes in the effective scattering length and
effective slope parameters are the key elements of our microscopic
approach to the kaonic atom level shifts.

\section{Non-local optical potential: phenomenology}

In this section, in order to make an estimate of non-local effects in
kaonic atoms, we calculate the spectra with an optical
potential $U_{\rm opt.}(r,\vec \nabla) $ deduced from the kaon self
energy (\ref{beff}) but use a phenomenological Ansatz for its
non-local structure. Our starting point is the Klein-Gordon equation
\begin{equation}
\vec \nabla^2 \phi(r) + 
\left[ \left(\mu-E-{i\, \Gamma \over 2}-V_{\rm e.m.}(r) \right)^2 - \mu^2 \right] \phi(r)
= 2\, \mu \,U_{\rm opt.}(r,\vec \nabla) \,\phi(r),
\label{kg}
\end{equation}
where $E$ and $\Gamma$ are the binding energy and width of the kaonic
atom, whereas $\mu $ is the reduced kaon mass in the $K^-$ nucleus
system. The electromagnetic potential $V_{\rm e.m.}$ is the sum of the
Coulomb potential and the Uehling and K\"allen-Sabry vacuum
polarization potentials \cite{Blomqvist} folded with the nucleus
density profile.  The nuclear densities are taken from
\cite{Friedman}, where they are obtained by unfolding a gaussian
proton charge distribution from the tabulated nuclear charge
distributions \cite{Vries}. We solve the Klein-Gordon equation
(\ref{kg}) using the computational procedure of Krell and Ericson
\cite{Krell}.

For the optical potential $U_{\rm opt.}(r,\vec \nabla )$, appearing on
the right-hand-side of (\ref{kg}), we make the following Ansatz:
\begin{eqnarray}
U_{\rm opt.}^{(i)} = U_{\rm opt.}^{(0)} &+&  V_i\;, \quad \quad 
2 \mu \,U_{\rm opt.}^{(0)}  = - 4 \pi \left(1+{m_K \over m_N} \right)
a\left[\rho(r)\right]  \rho(r)\; ,
\nonumber\\ \nonumber \\
2 \mu \,V_1  &=&  4 \pi \left(1+{m_K \over m_N} \right)
b[\rho(r)] \rho(r)  \vec \nabla^2  \;,
\nonumber\\
2 \mu \,V_2  &=&  4 \pi \left(1+{m_K \over m_N} \right)
b[\rho(r)] \vec \nabla \rho(r) \cdot \vec \nabla \; ,
\nonumber\\
2 \mu \,V_3  &=&  4 \pi \left(1+{m_K \over m_N} \right)
b[\rho(r)] [\vec \nabla^2 \rho(r)] \;.
\label{V}
\end{eqnarray}
In contrast to the approach of \cite{Toki} we do not fit the
spectrum. The effective scattering length $a[\rho]$ and the effective
slope parameter $b[\rho]$ in (\ref{V}) are determined by the expansion
of the $K^-$ self energy at small momenta (see (\ref{beff})): we
identify $a= a_{\rm eff.}$ and $b= b_{\rm eff.}/k_F$.  The optical
potentials (\ref{V}) follow from (\ref{beff}) with ${\vec q}\,$
replaced by the momentum operator $-i\vec \nabla$. Of course, this
heuristic procedure is not unique and may lead to different ways of
ordering the gradients. For this reason we study the three different
cases in (7) separately. Although our procedure of constructing the optical
potential is not strict, it allows us to make an estimate of the
magnitude of the non-local effects, usually neglected in other
approaches. A more systematic derivation of the non-local part of the
optical potential is presented in the subsequent section.

Here we wish to emphasize an important aspect related to the expansion
suggested in (4).  In the previous section we argued that the typical
kaon momentum $q$ up which (4) should represent the kaon self energy
is rather small with $q \sim 20 $ MeV. This observation is evidently
true if the gradient in (7) is acting on the kaon wave function.
However, if the proper gradient ordering asks for a sizeable
contribution in which the gradient is acting on the nucleus density
profile this cannot be true anymore.  Such terms probe the surface
thickness of the nucleus which in turn would require that the
expansion in (4) represents the kaon self energy up to much larger
momenta $q \sim 200 $ MeV. This leads to a severe conflict, because
the expansion in (4) can only be performed for the typical momentum
$q$ smaller than the typical Fermi momentum: if the non-local Pauli
blocking contribution is to be expanded one must either assume $q <
k_F$ or $q > k_F$ (see Appendix). We conclude that given the effective
functions $a_{\rm eff.}$ and $b_{\rm eff.}$ only, we can at most
expect to derive terms in the optical potential linear in $\nabla \rho
$. A term like given in the last line of (7) is outside the scope of
this work, because its systematic derivation requires a more general
input. Note that an analogous consideration applies to the expansion
of the self energy in powers of $\omega -m_K $.

The nuclear energy shift $\Delta E$ and the width $\Gamma$ of the
3d$\to$2p transition (${}^{10}_{\,\,5}$B,${}^{12}_{\,\,6}$C) and the
4f$\to$3d transition (${}^{27}_{13}$Al,${}^{32}_{16}$S) are presented
in Tab. 1. In the second column (LDT) we recall the results obtained
with the optical potential determined by the low-density theorem
(\ref{LDT}).  In the next column we present the results obtained with
the effective scattering length $a=a_{\rm eff.}$, as shown in
Fig. 2. The use of the density dependent scattering length improves
the agreement with the empirical data as compared to the density
independent scattering length.  In particular the level widths
increase towards the empirical values.  In the next three columns we
show the results obtained with the non-local potentials of (\ref{V})
with $b=b_{\rm eff.}/k_F$. For all considered nuclei we observe the
same type of the change in the spectrum. In contrast to $U_{\rm opt.}^{(1)}$, 
the effects of $U_{\rm opt.}^{(2)}$  are large. Whereas the level energies  
are affected only moderately by $U_{\rm opt.}^{(2)}$ the widths of the kaonic 
atom levels are increased significantly by hundreds of eV. This is an effect 
leading towards the proper spectrum. However, in our case the strength of $b$ is 
too small to obtain a completely successful agreement with data though 
a semi-quantitative description is achieved. An improved phenomenological description 
of the selected cases follows if the effective scattering length $a_{\rm
eff.}$ of Ramos and Oset \cite{Ramos} is combined with the effective
slope parameter $b_{\rm eff.}$ of Fig. 2. The results, shown in the
second last row of Tab. 1, agree remarkably well with the data. Of
course it is not consistent to combine the two different calculations
\cite{ML,Ramos}. The last column is included only to demonstrate that
there may be many different ways to describe the K-atom data in a
phenomenological approach.

From our previous discussion it should not be a surprise that $U_{\rm opt.}^{(3)}$ 
leads to unphysical results as clearly proven in Tab. 1. The huge effects found 
for that form of the optical potential are an indication that one should also 
investigate non-local effects involving terms like $\nabla^2 \rho $ or $ (\nabla \rho )^2 $
systematically. Such effects are outside the scope of the present study because 
they are clearly not determined by the effective slope parameter $b_{\rm eff.}$. 
They will be carefully investigated in a separate work.

\savebox{\tabela}{\vbox{
\tabcolsep=1.8mm
\begin{tabular}{|c|c|r|r|r|r|r|r|r|}
\hline
\multicolumn{2}{|c|}{Nucleus} &  LDT &$U^{(0)}_{\rm opt.}$&
$U^{(1)}_{\rm opt.}$&$U^{(2)}_{\rm opt.}$& $U^{(3)}_{\rm opt.}$& RO+L & experiment \\
\hline 
\hline
${}^{10}_{\,\,5}$B  & $-\Delta E $   & 174 & 222 &  238    &  236    &  -128     & 237 & 208 $\pm$ 35 \\
\cline{2-9}
             & $\Gamma$  & 322 & 441 &  405    &  569    &  11     & 713 & 810 $\pm$ 100 \\
\hline
\hline
${}^{12}_{\,\,6}$C  & $-\Delta E $   & 475 & 603 &  638    &  638    & -218     & 674 &  590 $\pm$ 80 \\
\cline{2-9}
          & $\Gamma$      & 761 &1022 &  926    & 1342    &  -38     &1694 & 1730 $\pm$ 150 \\
\hline   
\hline
${}^{27}_{13}$Al    & $-\Delta E $ &  92   &  115  &   130  &  141   & -128  &  135  &  130 $\pm$ 50 \\
\cline{2-9}
        & $\Gamma$    & 220   &  293  &   282  &  395   & -52  &  478  &  490 $\pm$ 160  \\
\hline       
\hline
${}^{32}_{16}$S     & $-\Delta E $  & 497 & 625 &   683   & 729     &   -378   & 748  &  550 $\pm$ 60 \\
\cline{2-9}    
      & $\Gamma$     &1023 &1324 &  1264   &1748     &   -41   & 2105  & 2330 $\pm$ 200 \\
\hline       
\end{tabular}}}
\begin{table}[t]
\vspace{1.0cm}
\usebox{\tabela}
\caption{Collection of our results for different nuclei. The energy level shifts
$\Delta E$ and widths are all given in eV. LDT denotes the results
obtained with the use of the low-density theorem with
$a_{KN}=(-0.18+i\,0.67)$ fm.  The last column gives the empirical
value of \cite{backenstoss72} (for ${}^{10}_{\,\,5}$B, ${}^{12}_{\,\,6}$C and
${}^{32}_{16}$S) and \cite{barnes} (for ${}^{27}_{13}$Al).  The results
obtained with the optical potentials (\ref{V}) are listed in the
columns denoted by $U^{(0)}_{\rm opt.}$, $U^{(1)}_{\rm opt.}$,
$U^{(2)}_{\rm opt.}$ and $U^{(3)}_{\rm opt.}$, respectively. RO+L is
the result obtained with $a[\rho]$ given by Ramos and Oset
\cite{Ramos}, and $b[\rho]$ given in \cite{ML}.}
\label{tab:results}
\end{table}

We stress that we do not advocate that either of the three
phenomenological gradient orderings is realistic. Tab. 1 is included
in this work to demonstrate that non-local effects are important and
in particular that it is crucial to derive the proper gradient
ordering systematically. Furthermore we point out that the expansion
of the kaon self energy in small momenta $q$ in (\ref{beff}) requires
that the contribution of the region $0< k_F < q \simeq 20$ MeV to the
kaonic level shifts is insignificant. In order to verify this
assumption we artificially modified the slope function such that it is
zero at $k_F=0$ but agrees to good accuracy with $b_{\rm eff.}(k_F)$
for $k_F>50$ MeV. This is achieved with:
\begin{eqnarray}
b_{\rm eff.}(k_F) \to \sqrt{\frac{k_F^2}{(20\,{\rm MeV})^2+k_F^2}}\; b_{\rm eff.}(k_F) \;.
\label{beff-rescale}
\end{eqnarray}
We considered the implication of the modified slope function
(\ref{beff-rescale}) for the level shift of the $^{12}_{\;6}C\,K^-$
system. The optical potentials $U^{(1)}_{\rm opt.}$ and
$U^{(2)}_{\rm opt.}$ are found rather insensitive to the modification
of $b_{\rm eff.}(k_F)$ according to (\ref{beff-rescale}), the level
shift and width change by less than $2 \%$.

It is instructive to compare our results with the analysis of Ramos
and Oset \cite{Ramos}.  In their approach the real part $a[\rho]$
becomes positive at somewhat smaller densities, as compared to the
calculation \cite{ML}. A more rapid change of Re $a[\rho]$ leads to a
stronger attraction of the optical potential $U_{\rm opt.}^{(0)}$ and
better agreement with empirical level shifts \cite{baca}\footnote{Note
that in \cite{Okumura} an unrealistic density profile is used for
light nuclei.  If the effective scattering length of \cite{Ramos} and
the density profile (\ref{profile}) is used for the $^{12}_{\;6}$C
K$^{-}$ system the nuclear level shift is $\Delta E = -623$ eV and
$\Gamma= 1290$ eV.}.  We point out that the recent calculations
\cite{baca,Okumura} are not conclusive since first, they ignore the
important non-local structure of the optical potential and second, the
optical potential is still subject to sizeable uncertainties from the
microscopic kaon-nucleon interaction. Strong non-local effects are an
immediate consequence of the $\Lambda(1405)$-resonance structure and
an important part of any microscopic description of kaonic atom
data. The uncertainties in the optical potential reflect the poorly
determined subthreshold kaon-nucleon scattering amplitudes. The
isospin zero amplitude of \cite{Ramos} and \cite{ML} differ by almost
a factor two in the vicinity of the $\Lambda(1405)$-resonance. Naively
one might expect that the amplitudes of \cite{ML} are more reliable
since they are constructed to reproduce the amplitudes of \cite{Waas}
which are based on a chiral $SU(3)$ analysis of the kaon-nucleon
scattering data to chiral order $Q^2$ as compared to the analysis of
Ramos and Oset, which considers only the leading order $Q$. On the
other hand the amplitudes of \cite{Ramos} include further channels
implied by $SU(3)$-symmetry but not included in the amplitudes of
\cite{Waas}. Also the work by Ramos and Oset includes the effect of an
in-medium modified pion propagation not included in \cite{ML}. Note,
however that such effects are not too large at small density. Clearly
a reanalysis of the scattering data and the self consistent many-body
approach in an improved chiral $SU(3)$ scheme is highly desirable
\cite{Kolomeitsev}.

\section{Semi-classical expansion}

In this section we suggest to systematically construct the non-local part of the optical 
potential by a  semi-classical expansion of the nucleus structure. In the previous section 
we have demonstrated that the ordering of the gradient terms has dramatic influence on the 
kaonic level shifts.  Naively one may conjecture that the gradient ordering should be 
s-wave like since the vacuum kaon-nucleon interaction is s-wave dominated. A perturbative 
s-wave interaction would give rise to the gradient ordering suggested in \cite{ML} in contrast
to the phenomenological successful Kisslinger potential  \cite{Kisslinger} employed in \cite{Toki}.
However, since the kaon self energy is obtained in a non-perturbative 
many-body approach it is not obvious as to which gradient ordering to choose. In particular
it is unclear to what extent the gradients are supposed to act on the function $b_{\rm eff.}(k_F)$
in (\ref{beff}). Since $b_{\rm eff.}(k_F)$ is a rapidly varying function in the nuclear density 
such effects have to be considered.

The starting point of our semi-classical  approach is the Klein-Gordon equation 
\begin{eqnarray}
\Big[ \vec \nabla^2 -\mu^2+ \Big( \omega-V_{\rm e.m.}(r) \Big)^2
\Big] \phi({\vec r}\,)=
2\,\mu\,\int d^3r'\, U_{\rm opt.} (\omega ,{\vec r},{\vec 
r}\,') \,\phi({\vec r}\,') 
\end{eqnarray}
with the non-local $K^-$ nuclear optical potential $U_{\rm opt.} (\vec r,\vec r\,' )$ and
$\omega = \mu -E-{i\, \Gamma \over 2}$.  
Since the binding energy $E$ of a kaonic atom is of the order of 
hundreds of keV it will be justified to expand the nuclear optical potential in the kaon 
energy $\omega $ around the reduced kaon mass $\mu$. Here we exploit the obvious fact 
that the nuclear part of the potential varies extremely smoothly at that scale. 

In order to derive the form of the non-locality in the nuclear 
optical potential it is useful to consider the in-medium kaon-nucleon scattering process. 
The  s-wave $K^-$-nucleon interaction of \cite{ML} implies that the scattering amplitude, 
$T_{KN}(\omega , \vec q\, ) $, depends exclusively on the sum of initial kaon  and 
nucleon momenta $\vec q $, and energies $\omega $. Close to the 
kaon-nucleon threshold the former can therefore be represented by 
the density dependent coefficients $a_{KN}(\rho), b_{KN}(\rho) $  
and $c_{KN}(\rho)$ 
\begin{eqnarray}
T_{KN}(\omega , \vec q \,; \rho ) &&= 4\,\pi\, 
\left(1+\frac{m_K}{m_N}\right) 
\Bigg(  a_{KN}(\rho)
+b_{KN}(\rho)\,\vec q\,^2
\nonumber\\
&&\;\;\;\;\;\;\;\;\;\;\;\;\;\;\;\;\;-c_{KN}(\rho )
\,\Big(\omega-m_N-m_K \Big) +\cdots 
\Bigg) \;,
\label{amplitude}
\end{eqnarray}
where we suppress higher order terms for simplicity. We note that 
at vanishing baryon density one expects from covariance 
$2\,(m_N+m_K)\, b_{KN}= c_{KN}$. In the nuclear medium, however, 
the functions $b_{KN}(\rho )$ and $c_{KN}(\rho ) $ are independent 
quantities. The model independent low-density characteristic is derived
in the Appendix: 
\begin{eqnarray} 
 b_{KN}(\rho)  =  -\frac{1}{6\,\pi}\,
\frac{3\,\kappa-1}{(1+\kappa)^2 }\,
\frac{1}{k_F} \,
\Bigg( \left(a_{KN}^{(I=0)}\right)^2 + 
3\left(a_{KN}^{(I=1)}\right)^2 \Bigg) +{\mathcal O}\left( k_F^0\right),
\nonumber\\
 c_{KN} (\rho ) = 
-\frac{1}{\pi}\,\frac{m_K}{1+\kappa}\,\frac{1}{k_F}\, 
\Bigg( \left(a_{KN}^{(I=0)}\right)^2 + 
3\left(a_{KN}^{(I=1)}\right)^2 \Bigg) +{\mathcal O}\left( k_F^0\right).
\label{t-expand}
\end{eqnarray}
It should not come as a surprise that we find $b_{KN} \sim 1/k_F$ 
and $ c_{KN} \sim 1/k_F$ at small density. The leading term 
follows upon expansion of the scattering amplitude in powers of small momenta 
followed by the low density expansion. For details we refer to the Appendix. In fact 
the expansion in (\ref{amplitude}) stops to converge at $k_F< q$. Since the scattering 
amplitude leads to the kaon self energy via 
\begin{eqnarray}
\Pi_{K}(\omega , {\vec q}\,) &=& -4\,\int_0^{k_F}\frac{d^3p}{(2\pi)^3}
\,T_{KN}
\left( \omega +m_N+\frac{{\vec p}\,^2}{2\,m_N},\, {\vec q}+{\vec p} \right) \,,
\label{selfenergy}
\end{eqnarray}
we conclude that at small density (with $k_F < 50 $ MeV ) an accurate representation of 
the kaon self energy requires an infinite number of terms in (\ref{amplitude}). A 
manifestation of this fact is that $c_{\rm eff.} \neq k_F \, c_{KN} $ and 
$b_{\rm eff.}  \neq k_F \, b_{KN}  $ at small density. 
We observe that at somewhat larger density with $50 $ 
MeV $< k_F < 300 $ MeV, one  obtains a faithful representation of the  
self energy keeping only a few terms in (\ref{amplitude}). In any case  
our final optical potential (23) will be expressed directly in terms of 
$a_{\rm eff.}(k_F)$,$b_{\rm eff.}(k_F)$ and 
$c_{\rm eff.}(k_F)$ defined in (\ref{beff}) and extracted numerically form the kaon 
self energy of \cite{ML} (see discussion below (\ref{atilde})). 

In order to proceed and derive the non-local kaon optical potential 
we generalize (\ref{selfenergy}), which holds for infinite 
homogeneous nuclear matter, to the non-local system of a nucleus 
(see e.g. \cite{Thies}): 
\begin{eqnarray}
\!\!\!\!\!\!2\,\mu\,U_{\rm opt.}(\omega , {\vec r},{ \vec r}\,') &=&
- \sum_{n}\,\chi_n^{\dagger }({\vec r} \,) \,
T\Big(\omega-V_{\rm e.m.}({\vec r}\,')+E_n, {\vec r},{\vec r}\,'\, \Big)\,
\chi_n({\vec r}\,').
\label{non-local}
\end{eqnarray}
Here we introduced the non-local amplitude $T(\omega, 
\vec r,\vec r\,')$ and the nucleon wave function $\chi_n(\vec r \,) $ 
for a given nucleus in a shell model description. $E_n$ is the 
energy of the single particle wave function and the index $n$ sums 
all occupied shell model states of the nucleus. 

To make contact with the formalism for homogeneous nuclear matter 
it is useful to transform the amplitude $T(\omega , \vec q ;\vec R \,) $ to momentum space 
\begin{eqnarray}
T(\omega, {\vec r},{\vec r}\,') &=&
\int \frac{d^3q }{(2\pi)^3}\,e^{\,i\,{\vec q} \cdot ({\vec r}-{\vec r}\,')}
\,T(\omega, {\vec q}\,;{\vec R} \,)
\end{eqnarray} 
with $\vec R=(\vec r+\vec r\,')/2 $. In the limit of infinite and 
homogeneous nuclear matter the amplitude $T(\omega, \vec q ;\vec R 
)$ does not depend on $\vec R$ and equals the previously analyzed 
amplitude of (\ref{amplitude}). For the non-local system we 
identify: 
\begin{eqnarray}
T_{KN}(\omega , {\vec q}; \rho({\vec R}) ) =T(\omega, {\vec q} 
;{\vec R} ) 
\end{eqnarray}
with $\rho(\vec R) $ the density profile of the nucleus. 
We emphasize that the identification (15) neglects an explicit dependence of the 
in-medium scattering amplitude on $\nabla^2 \rho $. This is of no further concern 
here, because such effects are not addressed in this work.

We now aim at a shell model independent representation of the 
non-local optical potential by rewriting it in terms of  the 
nuclear density profile of the nucleus. Consider first 
the term  proportional to $ b_{KN}\,\vec q\,^2 $ in 
(\ref{amplitude}). We rewrite its contribution to the optical 
potential by  means of the semi-classical  expansion 
\begin{eqnarray}
&&b_{KN}({\vec R}\,)\,\sum_n \,\chi_n^\dagger({\vec r} \,)\, 
\chi_n({\vec r}\,')\, \vec \nabla_{{\vec r}\,'}\cdot \vec \nabla_{{\vec r}\,'} 
\,\delta^3({\vec r}-{\vec r}\,')
\nonumber\\
&=&\Bigg(
\frac{1}{4}\,\Delta \,\rho(r)\, b_{KN}(r)
-\frac{3}{5}\, b_{KN}(r)\,k_F^2(r)\,\rho(r) 
-\frac{1}{12}\,b_{KN}(r) \,\Delta\,\rho(r)
\nonumber\\
&-& b_{KN}(r)\,\frac{\Big(\vec \nabla 
\rho(r)\Big)^2}{36\,\rho(r)} 
+\vec  \nabla_{{\vec r }}\,b_{KN}(r)\,\rho(r)\,\cdot \vec \nabla_{{\vec r}}
\Bigg) \,\delta^3({\vec r}-{\vec r}\,')
+{\mathcal O}\left(\hbar^4\right),
\end{eqnarray}
where we did not include a spin-orbit force for simplicity (see 
e.g. \cite{Brack}). We emphasize that we have to drop all terms in (16) which 
involve more than one gradient acting on the density profile. For such terms 
the expansion in (4) does not make any sense.  On the other hand, the last 
term in (16) is well established and leads to a microscopic interpretation 
of the gradient ordering for $V_2$ in (7). We turn to $ c_{KN} $ in (\ref{amplitude}). 
This term appears most susceptible to shell effects since it probes 
the shell model energies $E_n$ explicitly. Again we perform the 
semi-classical  expansion 
\begin{eqnarray}
\sum_n \,E_n\,\chi_n^\dagger(\vec r\, )\,\chi_n(\vec r \,)
&=&\left( m_N+\bar \mu -\frac{\Delta }{18\,m_N}
-\frac{k_F^2(r)}{5\,m_N}\right) \rho(r) 
+{\mathcal O}\left( \hbar^4 
\right) \nonumber \\
\label{timederivative}
\end{eqnarray}
with $\bar \mu \simeq -8 $ MeV the one-nucleon separation energy of 
the nucleus.  Again one needs to realize 
that for instance the term with $\bar \mu -k_F^2/(5 m_N)$ in (17) is not determined 
reliably and therefore must be dropped. This follows, because the 
expansion of the self energy in powers of $\omega -m_K$ is applicable 
only for $\omega -m_K \sim 1$ MeV of the order of the $K$-atom binding energy.

We summarize the results of the semi-classical expansion. It leads to a 
description of $K^-$ atoms in terms of the empirical nuclear density profile. 
Upon collecting the various contributions we present the optical potential as 
implied by (\ref{amplitude}): 
\begin{eqnarray}
2\,\mu\,U_{\rm opt.} (\omega,\vec r, \vec \nabla ) &=&
-4\,\pi \left(1+\frac{m_K}{m_N} \right)
\Big(\bar a_{KN}[\rho(r)]\,\rho(r)+s_{KN}[\rho(r)] \Big)
\nonumber\\
&+&4\,\pi \left(1+\frac{m_K}{m_N} \right)
\vec \nabla \,b_{KN} [\rho (r)]\,\rho(r)\,\cdot \vec \nabla
\nonumber\\
&+& 4\,\pi \left(1+\frac{m_K}{m_N} \right)
\Big(\omega -V_{\rm e.m.}(r)-m_K \Big)c_{KN}[\rho(r)]\, \rho(r),
\nonumber \\
\label{koptical}
\end{eqnarray}
where we represent the non-local optical potential, $U_{\rm opt.}({\vec 
r}, {\vec r}\,')$, in terms of the differential operator, 
$U_{\rm opt.}({\vec r},\vec \nabla )$, with: 
\begin{eqnarray}
U_{\rm opt.}(\omega,{\vec r},\vec \nabla )\,\phi({\vec r}\,) =
\int d^3r'\, U_{\rm opt.} (\omega ,{\vec r},{\vec r}\,') \,\phi({\vec 
r}\,')  \;.
\end{eqnarray}
We further introduce the renormalized effective scattering 
length $\bar  a_{KN}[\rho ]$ and a term $s_{KN}[\rho ]$ responsible 
for binding and surface effects
\begin{eqnarray}
\bar a_{KN}[\rho ] &=&a_{KN}(\rho )+\frac{3}{5}\,k_F^2\,
\left( b_{KN}(\rho )-\frac{1}{2\,m_N}\,c_{KN}(\rho ) \right),
\nonumber\\
s_{KN}[\rho ] &=&
\frac{1}{12}\, b_{KN}(\rho )
\,\left(\Delta \,\rho +\frac{\big(\vec \nabla \rho \big)^2}{3\,\rho} \right)
-\frac{1}{4}\,\Delta \, b_{KN}(\rho )\,\rho 
\nonumber\\
&-& c_{KN}(\rho ) \left( {\bar \mu}
-\frac{k_F^2}{2\,m_N}-\frac{\Delta }{18\,m_N} \right) \rho.
\label{atilde}
\end{eqnarray}
Equation (\ref{atilde}) is instructive since it suggests to 
identify $\bar a_{KN}(\rho) $ with the effective scattering length 
$a_{\rm eff.}(k_F ) $ introduced when analyzing the kaon self energy. 
It also gives a clear separation of effects we have under control from those 
which are beyond the scope of this work. According to our previous discussions all 
terms collected in $s_{KN}$ should be dropped. We arrive at our final form 
of the optical potential  by replacing 
\begin{eqnarray}
&& \bar a_{KN}(\rho ) \rightarrow  a_{\rm eff.}(k_F ) \,, \qquad  
b_{KN}(k_F ) \rightarrow  b_{\rm eff.}(k_F )/k_F \;, 
\nonumber\\
&& c_{KN}(\rho ) \rightarrow  c_{\rm eff.}(k_F )/k_F  \,,
\qquad  s_{KN}(\rho ) \rightarrow 0 \,,
\label{replacement}
\end{eqnarray}
in (\ref{koptical}) with the functions $a_{\rm eff.}(k_F), b_{\rm eff.}(k_F)$ 
and $c_{\rm eff.}(k_F)$ as extracted from  the kaon self energy of the self 
consistent calculation in \cite{ML}. We point out that this procedure 
recovers higher order terms in the expanded amplitude
(\ref{amplitude}). Note for instance that (\ref{atilde}) is written in such a 
way that $\bar a_{KN}$ includes the trivial renormalization of 
the scattering length implied by (\ref{selfenergy}). This justifies the identification 
of $\bar a_{KN}$ with $a_{\rm eff.}(k_F)$, because it avoids the double inclusion 
of such effects.

We present the final radial differential equation for  the kaon 
wave function $\phi({\vec r}\,) =\chi(r)\,Y_{lm}( 
\theta , \phi )/r $ as it follows from (\ref{koptical},\ref{atilde}) 
and (\ref{replacement}): 
\begin{eqnarray}
\!\!\!\!\!\!\!Z(r) \,\left(\frac{d^2}{d\,r^2}-\frac{l
(l+1)}{r^2}\right)\chi(r)&=& \bar Z(r)\,\Bigg[\mu^2- \Big( 
\mu-E-{\textstyle{i\over 2}}\,\Gamma-V_{\rm e.m.}(r)
\Big)^2\Bigg]\,\chi(r)
\nonumber\\
&+& 2\,\mu\,V(r)\,\chi(r)+ 2\,\mu\,S(r)\,
\left( \frac{d}{d\,r}\,\chi(r)\right)
\label{eff-KG}
\end{eqnarray}
with 
\begin{eqnarray}
2\,\mu\,V(r) &=&-\frac{8}{3\pi} \left( 1+\frac{m_K}{m_N} \right)
\Bigg[ a_{\rm eff.}\left(k_F(r)\right) k^3_F(r)
+\frac{d}{r\,d\,r} \,b_{\rm eff.}\left(k_F(r)\right)k_F^2(r) \Bigg]\,,
\nonumber\\
Z(r) &=& 1-\frac{8}{3\pi } \left( 1+\frac{m_K}{m_N}
\right)k^2_F(r)\,b_{\rm eff.}\left(k_F(r)\right),
\nonumber\\
2\,\mu\,S(r) &=&\frac{8}{3\pi }\left( 1+\frac{m_K}{m_N} \right)
\frac{d}{d\,r}\,k_F^2(r)\,b_{\rm eff.}\left(k_F(r)\right),
\nonumber\\
\bar Z (r) &=& 1-\frac{8}{3\pi } \left( 1+\frac{m_K}{m_N}
\right)\frac{k^2_F(r)}{2\,m_K}\,c_{\rm eff.}\left(k_F(r)\right).
\label{eff-pot}
\end{eqnarray}
The effective Klein-Gordon equation (\ref{eff-KG}) shows a wave
function renormalization $Z(r)$, a mass renormalization $\bar Z(r)$
and a surface interaction strength $S(r)$. Note that in (\ref{eff-KG}) we approximated  
the linear energy dependence of (\ref{koptical}) by a quadratic one.

\begin{figure}[b]
\epsfysize=8.5cm
\begin{center}
\mbox{\epsfbox{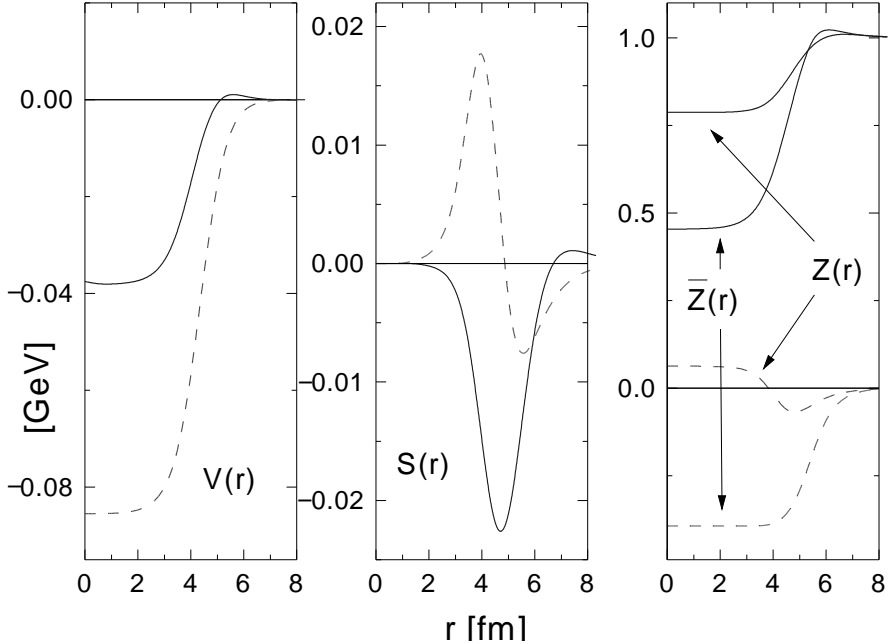}}
\end{center}
\caption{The effective potential $V(r)$, the effective surface function 
$S(r)$ and the effective renormalization functions 
$Z(r),\,\bar Z(r)$ as introduced in (\ref{eff-pot}) for the $^{58}_{28}$Ni K$^-$ system. 
The solid and dashed lines represent the real and imaginary parts 
respectively.} 
\label{fig3}
\end{figure}

We discuss the effect of the semi-classical optical potential
(\ref{eff-KG}) at hand of a $K^-$ bound at the $^{12}_{\;6}$C-core and
$^{58}_{28}$Ni-core. We use the unfolded density profiles of
\cite{Vries,Friedman}:
\begin{eqnarray}
&&\rho_{[^{12}_{\;6}C]}(r) = \rho_0 
\left(1+2.234\,\left(\frac{r}{1.516\,{\rm fm}}\right)^2 \right)
\exp \left( -\left(\frac{r}{1.516\,{\rm fm}}\right)^2\right) \;, 
\nonumber\\
&&\rho_{[^{58}_{28}{\rm Ni}]}(r) = \rho_0 
\left(1+\exp \left(\frac{r-4.134\, {\rm fm}}{0.500\,{\rm fm}} \right) \right)^{-1} \;,
\label{profile}
\end{eqnarray} 
with $\rho_0 $ determined by $4\,\pi\,\int d r\,r^2 \,\rho(r) =12$ for
carbon and $4\,\pi\,\int d r\,r^2 \,\rho(r) =58$ for nickel.  In
Fig. 3 the effective potential $V(r)$, the effective surface function
$S(r)$ and the renormalization functions $Z(r),\bar Z(r)$ defined in
(\ref{eff-pot}) are shown for the $^{58}_{28}$Ni density profile of
(\ref{profile}).  We point out that for the wave function and mass
renormalization functions $Z(r)$ and $\bar Z(r)$ we find quite large
deviations of about 50$\%$ from the asymptotic value 1. Also the
function $S(r)$ shows sizeable strength close to the nucleus surface.

We introduce the nuclear level shift, $\Delta E$, in terms
of the measured transition energy, $\Delta E_{\rm emp.} =\Delta E_{\rm
e.m.} +\Delta E $, and the theoretical electromagnetic
transition energy, $\Delta E_{\rm e.m.}$. For the $^{12}_{\;6}C\,K^-$
system we consider the 3d$\to$2p transition with $E_{\rm e.m.}$(2p) =
0.113763 MeV and $E_{\rm e.m.}$(3d) = 0.050458 MeV.  For the
$^{58}_{28}$Ni K$^-$ system we consider the 5g$\to$4f transition with
$E_{\rm e.m.}$(4f) = 0.641572 MeV and $E_{\rm e.m.}$(5g) = 0.409951
MeV.  Here the Coulomb potential and the vacuum polarization
potentials \cite{Blomqvist} are folded with the nucleus density
profile (\ref{profile}).  We use $m_{K^-}=493.677$ MeV and
$1/\alpha =137. 035989$. The empirical transition energy is $\Delta
E_{\rm emp.}= 62.73\pm 0.08$ keV with $\Gamma = 1.73\pm 0.15$ keV
\cite{backenstoss72} for carbon and $\Delta E_{\rm emp.}= 231.408\pm
0.052$ keV with $\Gamma =1.230 \pm 0.140$ keV for nickel. Note that
the resulting empirical nuclear level shifts $\Delta E =
-573\pm 80$ keV for carbon and $\Delta E = -213 \pm 52$
keV for nickel differ slightly from the values given in
\cite{backenstoss72,batty79} and shown in Tab. 2. The reason for which
is an old value for the kaon mass, a slightly different nucleus
profile and the inclusion of further small correction terms.

\tabcolsep=1.2mm
\begin{table}{}
\begin{tabular}{|l||c|c||c|c|} \hline

& \multicolumn{2}{|c||}{ $^{12}_{\;6}$C (3d$\to$2p)}   & \multicolumn{2}{|c|}{ $^{58}_{28}$Ni (5g$\to $4f)} \\ 
\cline{2-5} 

& $-\Delta E$ [eV] & $\Gamma $ [eV] & $-\Delta E$ [eV] & $\Gamma $ [eV]
\\ \hline \hline 

$a_{KN}$  & 475 & 761 & 199 & 476 \\ \hline

$a^{(\rm eff.)}_{KN}$   & 593 & 1362  & 143 & 932
\\ 
\hline 

$b_{\rm eff.} = 0\;\; \& \;\;c_{\rm eff.}$ = 0 & 603 & 1022 & 248 & 627\\ \hline

$c_{\rm eff.}$ = 0 & 589 & 1109 & 278   & 759   \\ \hline

full\, theory      & 569 & 1098 &  233  & 701   \\ \hline

experiment \cite{backenstoss72,batty79} & 590 $\pm$ 80 & 1730 $\pm $ 150 & 246 $\pm 52$ & 1230 $\pm 140$\\ 
\hline 
\end{tabular}
\caption{Results for the  
$^{12}_{\;6}$C K$^-$ and $^{58}_{28}$Ni K$^-$ systems. The first two rows show 
the results from the optical potential 
(\ref{LDT}) with i) the empirical repulsive scattering length  $a_{KN} = 
(-0.18+i\,0.67)$ fm and ii) the effective attractive scattering length 
$a^{(\rm eff.)}_{KN} =(0.63+i\,0.89)$ fm. The remaining rows show the result with the semi-classical  
optical potential (\ref{eff-pot}) with  iii) $b_{\rm eff.}=0=c_{\rm eff.}$, iv)  
with $c_{\rm eff.} = 0$ and v) with $a_{\rm eff.}, b_{\rm eff.}$ and $c_{\rm eff.}$ from Fig. 2.}
\end{table}

In Tab. 2 we compare the results from the low-density optical potential 
(\ref{LDT}) with the result of the semi-classical optical potential.
The full semi-classical  potential represents the empirical level shifts 
quantitatively but underestimates the empirical widths systematically. 
As can be seen from Tab. 2 the non-local effects increase the level 
widths by only a small amount towards the empirical values. 
The results obtained with the semi-classical potential differ significantly from
the results obtained with phenomenological potential $V_2$ in (\ref{V}) with an ad-hoc 
gradient ordering. We emphasize that the crucial assumption of our approach, namely 
that the results are insensitive to the precise form of the slope functions 
$b_{\rm eff.}(k_F)$ and $c_{\rm eff.}(k_F)$ in the extreme surface region 
$k_F< q \simeq 20$ MeV, which justifies the expansion (\ref{beff}), is fulfilled. 
The size of the level shifts and widths, using $b_{\rm eff.}(k_F)$ and 
$c_{\rm eff.}(k_F)$ functions modified according to (\ref{beff-rescale}), are 
affected by less than $3 \%$ for the $^{12}_{\;6}$C K$^-$ and $^{58}_{28}$Ni K$^-$ 
systems. Note that this uncertainty is certainly not resolved by the accuracy of 
the available empirical level shifts and widths.

\section{Summary}

We summarize the findings of our work. The kaon self energy shows a
rapid density, energy and momentum dependence in nuclear matter. This
rich structure is a consequence of a proper many-body treatment of the
$\Lambda (1405)$ resonance structure in nuclear matter invalidating
any mean field type approach to kaon propagation in nuclear matter at
moderate densities. Thus non-local terms in the optical potential are
large and quantitatively important. If included phenomenologically in 
the calculation of the kaonic atom spectra they cause substantial 
additional shifts of the binding energies and level widths. Since the 
kaon self energy must be obtained in a non-perturbative many-body approach 
it is not immediate as to which gradient ordering to choose. A phenomenological
p-wave like ordering of gradient terms with the strength taken from
the many-body calculation of \cite{ML,Ramos} leads to a satisfactory
description of the kaonic atom level shifts.  However, a proper microscopic 
description of $K^-$-atom data requires a careful derivation of the non-local 
part of the optical potential. Given the kaon self energy as evaluated for 
infinite nuclear matter, one can at most establish non-local terms where a 
gradient is acting on the kaon wave function. This follows in particular, 
because the low-momentum expansion of the self energy converges only for 
$q< k_F$ once the important Pauli blocking effects are considered. In this 
work we established that for such non-local effects the typical kaon momentum 
is indeed smaller than the typical Fermi momentum as probed in K-atoms. This 
justifies an expansion of the kaon self energy in small momenta. 
We reiterate that given only the kaon self energy of infinite nuclear matter 
any contribution considered for the optical potential which involves $\nabla^2 \,\rho$ 
is not controlled and therefore must be dropped. 

We derived the microscopic structure of the optical potential by means of the 
semiclassical expansion in the second part of this work systematically. 
At present this approach reproduces the empirical level shifts quantitatively
but underestimates the empirical widths systematically. The microscopic optical 
potential leads to less pronounced non-local effects as compared to the 
phenomenological optical potential with p-wave like gradient ordering. Further 
improvements are conceivable. For instance the non-local effects involving 
$\nabla^2 \rho$ should be studied and also p-wave interaction terms should 
be considered. Moreover, the kaonic atom data appear very sensitive to the 
precise microscopic interaction of the kaon-nucleon system. An improved microscopic 
input for the many-body calculation of 
the kaon self energy is desirable. This is supported by the recent chiral $SU(3)$ 
analysis of the kaon-nucleon scattering data which includes for the first time 
also p-wave effects systematically \cite{Kolomeitsev}. It is found that the subthreshold
kaon-nucleon scattering amplitudes differ strongly from those of
\cite{Waas,Ramos} and that p-wave effects in the isospin one channel
are strong. An evaluation of the many-body effects based on the
improved chiral SU(3)-dynamics is in progress \cite{Korpa}.

\section*{Acknowledgments}

W.F. thanks the members of the Theory Group at GSI for very warm
hospitality.  M.L. acknowledges partial support by the Polish
Government Project (KBN) grant 2P03B00814.

\section{Appendix}

In this appendix we derive the contribution from Pauli blocking 
to the in-medium kaon-nucleon scattering process and the kaon self energy. 
Pauli blocking induces a model independent term in the scattering amplitude
$T^{(P.B.)}_{KN} (\omega+m_N ,\vec q\,)$ and self energy 
$\Pi^{(P.B.)}_{K} (\omega, \vec q\,)$ proportional to 
the squared s-wave scattering lengths:
\begin{eqnarray}
&&T^{(P.B.)}_{KN} (\omega+m_N ,\vec q \,) = \pi\left(1+\kappa\right) 
\left(\left(a_{KN}^{(0)}\right)^2
+3\,\left(a_{KN}^{(1)} \right)^2
\right) I(\omega, \vec q\, )\;,
\nonumber\\
&& \Pi^{(P.B.)}_{K} (\omega, \vec q) = 
-4\,\int_0^{k_F} \frac{d^3 p}{(2\pi)^3} 
\,T^{(P.B.)}_{K N} \big(\omega +m_N+p^2/(2\,m_N), \vec p+\vec q\,\big),
\nonumber \\
\label{}
\end{eqnarray}
with $\kappa = m_K/m_N$ and 
\begin{eqnarray}
&&\!\!\!\!\!\!I(\omega, \vec q\, ) = i\,p_{\rm cm}[\omega,\vec q\,]
\nonumber\\
&& -\int_0^{k_F}\frac{d^3l }{(2\pi)^3} 
\,\frac{4\,\pi \,\big(1+\kappa \big)}
{m_K^2+\Delta m_K^2- \omega^2+\big(1+\omega/m_N \big)
\,\vec l\,^2-2\,\vec l \cdot \vec q+\vec q\,^2 -i\,\epsilon },
\nonumber\\
\nonumber \\
&& \sqrt{(\omega+m_N)^2-\vec q\,^2} =\sqrt{m_N^2+p^2_{\rm cm}[\omega,\vec q\,] }
+\sqrt{m_K^2+\Delta m_K^2+p^2_{\rm cm}[\omega,\vec q\,]}.
\nonumber \\
\label{def-pauli}
\end{eqnarray}
Note that in the denominator of (\ref{def-pauli}) we neglected
relativistic correction terms of the form $(l^2)^{n+1}/m_N^{2 n}$ or
$\omega\,(l^2)^{n+1}/m_N^{2 n+1}$ with $n\geq 1$ but kept the
correction term $\omega /m_N$. The latter term is important since it
renders the derivatives $\partial_\omega \Pi (m_K,0)$ and
$\partial_{q^2} \Pi (m_K,0)$ finite and well defined at $\omega =m_K$
and $q^2=0$.  We emphasize that the inclusion of the kaon mass
modification $\Delta m_K^2 \sim k_F^3 $ as required in a
self-consistent calculation \cite{ML} leads to a well defined behavior
of $I(\omega, \vec q\,)$ close to $\omega \simeq m_K$ and $\vec
q\simeq 0$. Similarly, as suggested in \cite{Garcia}, one may include
nuclear binding effects in $I(\omega , \vec q\,)$ by replacing $m_N\to
m_N+\Delta m_N$ and $\omega \to \omega - \Delta m_N$ in the r.h.s of
(\ref{def-pauli}). This leads to identical results for the low-density
limits $\partial_\omega \,T^{} (m_K+m_N, 0)$ and $\partial_{q^2}
\,T^{} (m_K+m_N, 0)$ since only the combination $\Delta m_K +\Delta
m_N $ is active at leading order and will cancel identically. It is
straightforward to derive:

\begin{eqnarray}
\frac{\partial }{\partial \,\omega }\,I(m_K, 0) &=& \frac{4}{\pi}
\,\frac{m_K}{k_F} \,\frac{1}{1+\kappa}
+{\mathcal O}\left( k_F^0\right)\;,
\nonumber\\
\frac{\partial}{\partial \,q^2 }\,I(m_K, 0) &=& 
-\frac{2}{3\, \pi}\,\frac{1}{k_F}\,\frac{3\,\kappa-1}{(1+\kappa)^2}
+{\mathcal O}\left( k_F^0\right)\;,
\label{der-i}
\end{eqnarray}
leading to our result (\ref{t-expand}). Note that the two terms in $I(\omega , \vec q\,)$ 
lead to contributions proportional to $1/\sqrt{\Delta m_K+\Delta m_N}$ in (\ref{der-i}) 
which 
however cancel identically. Similarly one may expand 
\begin{eqnarray}
&&\int_0^{k_F} \frac{d^3p}{(2\pi)^3} \,I\big(m_K+p^2/(2\,m_N),\vec p\, \big)
= -\frac{k_F^4}{4\,\pi^3}\,
\frac{1-\kappa^2+\kappa^2\,\ln \kappa^2 }{(1-\kappa)^2\,(1+\kappa)}
+{\mathcal O}\left( k_F^5\right)\;,
\nonumber\\
&&\frac{d}{d\,\omega}\Bigg|_{\omega =m_K}\,
\int_0^{k_F} \frac{d^3p}{(2\pi)^3} \,I\big(\omega+p^2/(2\,m_N),\vec p\, \big)
=- \frac{k_F^2\,m_K}{2\,\pi^3}\,\frac{\ln | \kappa |}{1+\kappa } 
+{\mathcal O}\left(k_F^3 \right)\;,
\nonumber\\
&&\frac{d}{d\,q^2}\Bigg|_{q^2 =0}\,
\int_0^{k_F} \frac{d^3p}{(2\pi)^3} \,I\big(m_K+p^2/(2\,m_N),\vec p+\vec q\, \big)
= \frac{k_F^2}{4\,\pi^3}\Bigg( 
\frac{7}{3}\,\frac{\ln |\kappa |}{1+\kappa }
\nonumber\\
&& \qquad \qquad \qquad \qquad+ \frac{2}{3}\,
\frac{1-\kappa^2+2\,\kappa^2\,\ln | \kappa| }{(1+\kappa)\,(1-\kappa)^2}\Bigg)
+{\mathcal O}\left(k_F^3 \right)\;,
\label{int-limit}
\end{eqnarray}
and arrive at the low-density limit of $b_{\rm eff.}(0)$ and $c_{\rm eff.}(0)$ as given 
in (\ref{self-expand}). For completeness we included in (\ref{int-limit}) the integral 
leading to the $k_F^4$-term in the self energy given in \cite{ML}. Its limit for small 
$\kappa$ was derived in \cite{Oset}.


\begin{thebibliography}{9}

\bibitem{ML}
M. Lutz, Phys. Lett. {\bf B426}, 12 (1998); M.F.M. Lutz, in {\it 
Proc. Workshop on Astro-Hadron Physics}, Seoul, Korea, October, 1997, nucl-th/980233.

\bibitem{report}
E. Friedmann, A. Gal and C.J. Batty, {Phys. Rep.} {\bf 287}, 385 (1997).

\bibitem{Friedman}
E. Friedman, A. Gal and C.J. Batty, {Nucl. Phys.} {\bf A579}, 518 (1994).

\bibitem{Thies}
M. Thies, {Nucl. Phys.} {\bf A298}, 344 (1978).

\bibitem{Bardeen}
W.A. Bardeen and E.W. Torigoe, {Phys. Rev.} {\bf C3}, 1765 (1971).

\bibitem{Alberg}
M. Alberg, E.H. Henely and L. Wilets, {Ann. Phys.} {\bf 96}, 43 (1976).

\bibitem{Toki}
M. Mizoguchi, S. Hirenzaki and H. Toki, {Nucl. Phys.} {\bf A567}, 893 (1994); 
{Nucl. Phys. }{\bf A585}, 349c (1995).

\bibitem{Brockmann}
R. Brockmann and W. Weise, {Nucl. Phys.} {\bf A308}, 365 (1978).

\bibitem{Mares}
E. Friedman, A. Gal, J. Mares and A. Cieply, Phys. Rev. {\bf C60}, 024314 (1999).

\bibitem{Waas}
T. Waas, N. Kaiser and W. Weise, {Phys. Lett.} {\bf B365}, 12 (1996).

\bibitem{Koch}
V. Koch, {Phys. Lett. }{\bf B337}, 7 (1994); A. Ohnishi, Y. Nara
and V. Koch, Phys. Rev. {\bf C56}, 2767 (1997).

\bibitem{Blomqvist}
J. Blomqvist, { Nucl. Phys.} {\bf B48}, 95 (1972).

\bibitem{Vries}
H. de Vries, C.W. de Jaeger and C. Vries, Atomic Data and Nuclear 
Data {\bf 36}, 495 (1987).

\bibitem{Krell}
M. Krell and T.E.O. Ericson, {Jour. Comp. Phys.} {\bf 3}, 202 (1968). 



\bibitem{backenstoss72}
G. Backenstoss, A. Bamberger, I. Bergstr\"om, P. Bounin,
T. Bunaciu, J. Egger, S. Hultberg, H. Koch, M. Krell, U. Lynen,
H.G. Ritter, A. Schwitter, and R. Stearns, Phys. Lett. {\bf B38}, 181 (1972).

\bibitem{barnes}
P.D. Barnes, R.A. Eisenstein, W.C. Lam, J. Miller, R.B. Sutton,
M. Eckhause, J.R. Kane, R.E. Welsh, D.A. Jenkins, R.J. Powers, A.R. Kunselman,
R.P. Redwine and R.E. Segel, Nucl. Phys. {\bf A231}, 477 (1974).

\bibitem{Ramos}
A. Ramos and E. Oset, Nucl. Phys. {\bf A671}, 481 (2000);
we thank A. Ramos for providing us with a table of the effective scattering length. 


\bibitem{baca}
A. Baca, C. Garcia-Recio and J. Nieves, Nucl. Phys. {\bf A673}, 335 (2000).

\bibitem{Okumura}
S. Hirenzaki, Y. Okumura, H. Toki, E. Oset and A. Ramos, Phys. Rev. 
{\bf C61}, 055205 (2000);
we thank A. Ramos for communicating the above paper prior to publication.

\bibitem{Kolomeitsev}
M.F.M. Lutz and E.E. Kolomeitsev, in {\it Proceedings of the International Workshop XXVII
on Gross Properties of Nuclei and Nuclear Excitations}, Hirschegg, Austria, January, 2000.

\bibitem{Kisslinger}
L.S. Kisslinger {Phys. Rev.} {\bf 98}, 761 (1955).

\bibitem{Brack}
M. Brack, C. Guet and H.-B. Hakansson, Phys. Rep. {\bf 123}, 276 (1985). 

\bibitem{batty79}
C. J. Batty, S.F. Biagi, M. Blecher, S.D. Hoath, R.A.J. Riddle,
B.L. Roberts, J.D. Davies, G.J. Pyle, G.T.A. Squier, D.M. Asbury,
A.S. Clough, Nucl. Phys. {\bf A329}, 407 (1979).

\bibitem{Korpa}
C. Korpa and M. Lutz, in preparation.

\bibitem{Garcia}
C. Garcia-Recio, L.L. Salcedo and E. Oset, {Phys. Rev.} {\bf C39}, 595 (1989).

\bibitem{Oset}
C. Garcia-Recio, E. Oset, L.L. Salcedo, {Phys. Rev.} {\bf C37}, 194 (1988).

\end{thebibliography}
\end{document}